\documentclass[12pt,preprint]{aastex}
\shorttitle{Turbulent Model for Blazar Variability}
\shortauthors{A.~P. Marscher}

\begin{document}

\title{Turbulent, Extreme Multi-Zone Model for Simulating Flux and Polarization
Variability in Blazars}

\author{Alan P. Marscher}
\affil{Institute for Astrophysical Research, Boston University, 725 Commonwealth Avenue, Boston, MA 02215}
\email{marscher@bu.edu}

\begin{abstract}
The author presents a model for variability of the flux and polarization of blazars
in which turbulent plasma flowing at a relativistic speed down a jet crosses a 
standing conical shock. The shock compresses the plasma and accelerates electrons to
energies up to $\gamma_{\rm max} \gtrsim 10^4$ times their rest-mass energy, with
the value of $\gamma_{\rm max}$ determined by the direction of the magnetic field
relative to the shock front. The turbulence is
approximated in a computer code as many cells, each with a uniform magnetic field
whose direction is selected randomly. The density of high-energy electrons
in the plasma changes randomly with time in a manner consistent with 
the power spectral density of flux variations derived from observations of blazars.
The variations in flux and polarization are therefore caused by
continuous noise processes rather than by singular events such as explosive
injection of energy at the base of the jet. Sample simulations
illustrate the behavior of flux and linear polarization versus time that such a model produces.
The variations in $\gamma$-ray flux generated by the code are often, but not always,
correlated with those at lower frequencies, and many of the flares are sharply peaked.
The mean degree of polarization of
synchrotron radiation is higher and its time-scale of variability shorter toward
higher frequencies, while the polarization electric vector sometimes randomly executes
apparent rotations. The slope of the spectral energy distribution exhibits sharper breaks
than can arise solely from energy losses. All of these results correspond to
properties observed in blazars.
\end{abstract}

\section{Introduction}   

The blazar class of active galactic nuclei is characterized by extreme variability
of flux and polarization of nonthermal radiation across the electromagnetic spectrum. 
Statistical analyses of the flux variations --- in particular, the power-law
power spectra \citep{chat08,chat09,chat12,abdo11} --- imply that the variations are governed
by noise processes with higher amplitudes on longer time-scales. Furthermore,
the linear polarization of blazars ranges from a few to tens of percent and tends to
be highly variable in both degree and position angle \citep[e.g.,][]{darc07,mar10,itoh13}.
Although the range of degree of polarization and the tendency for the position
angle to be similar to or nearly transverse to the jet axis \citep[e.g.,][]{jor07}
can be explained with a helical magnetic field \citep{lyut05,push05}, the rapid variations
in polarization are not naturally reproduced in a model in which the field is
either 100\% globally ordered or completely chaotic on small scales. In the former
case, the ordered field provides stability to the polarization, while in the latter case
the polarization cancels in the presence of an essentially infinite number of random
field directions.

A natural explanation for the noise-like fluctuations of flux and polarization
in blazars is the presence of turbulence in the relativistic jets that emit the
nonthermal radiation. If one approximates the pattern of the turbulence
in terms of $N$ cells, each with a uniform magnetic field with random orientation,
the degree of linear polarization has a mean value of
$\langle \Pi \rangle \approx \Pi_{\rm max}N^{-1/2}$,
where $\Pi_{\rm max}$, typically between 0.7 and 0.8, corresponds to the uniform field
case \citep{Burn66}. If the cells pass into and out of the emission region, the
polarization will vary about the mean with a standard deviation
$\sigma_\Pi \approx 0.5\Pi_{\rm max}N^{-1/2}$ \citep{jones88}. The electric-vector
position angle (EVPA) $\chi$ of the polarization will also change with time in a random
manner that can often appear as a rotation over hundreds of degrees
\citep{jones88,mar08,darc09}.

The superposition of ordered and turbulent magnetic field components can provide
the combination of a preferred orientation of the polarization vector plus
fluctuations in $\Pi$ and $\chi$. Here the author introduces a numerical model
that creates this combination by passing turbulent cells through a standing shock
wave. \citet{dm87}, \citet{cc90}, \citet{gomez95}, and \citet{darc07} have proposed that a
stationary conical shock caused by a pressure mismatch with the external medium
corresponds to the bright compact structure (the ``core'') observed
at the upstream end of blazar jets imaged with very long baseline interferometry (VLBI)
at millimeter wavelengths. In this spectral range, relatively low optical depths
allow jets of many blazars to be viewed on sub-parsec scales with minimal obscuration by
synchrotron self-absorption.

The following two sections describe the physical model and a numerical code
that implements the model. This ``Turbulent Extreme Multi-Zone'' (TEMZ) code
calculates, as a function of time, the spectral energy distribution (SED)
from synchrotron radiation and inverse Compton (IC) scattering, as well as the linear
polarization of the synchrotron emission at various frequencies. Seed photons
for the scattering include infrared emission by hot dust in a parsec-scale
molecular torus, as well as synchrotron and IC photons
from a Mach disk (MD) on the jet axis. The MD, often called
a ``working surface'' when it occurs near the end of a jet, is a strong shock
oriented transverse to the jet axis that forms near the apex of a conical
shock when the jet maintains azimuthal symmetry \citep{cf76}.

After the description of the model in {\S}2, {\S}3 briefly reviews the tests carried out
to check the accuracy of the numerical code. Section 4 presents the results of
a small number of numerical simulations that serve as illustrations of
the variability patterns that the model generates, while {\S}5 presents the conclusions
of this work. A future study will include a more complete exploration of parameter space.

\section{Description of the TEMZ Model and Numerical Code}

The author has developed a numerical code designed to simulate the time-dependent SED from a 
relativistic jet in which turbulent plasma crosses a standing
conical shock plus Mach disk. (The code, originally written in the Fortran-77 computer language,
has also been translated to the C++ language by M. Valdez.) The run time on a common desktop
or laptop computer with a Linux operating system is hours or days, depending on the number of emission
zones (``cells''), values selected for the physical parameters, and the number of simulated time steps
involved (hundreds to thousands). The program calculates the SED from $1\times 10^{10}$ to
$5.6\times 10^{26}$ Hz, with four frequencies per decade, as well as the degree and position
angle of linear polarization (for synchrotron radiation only) at each time step.

\begin{figure}
\epsscale{.80}
\plotone{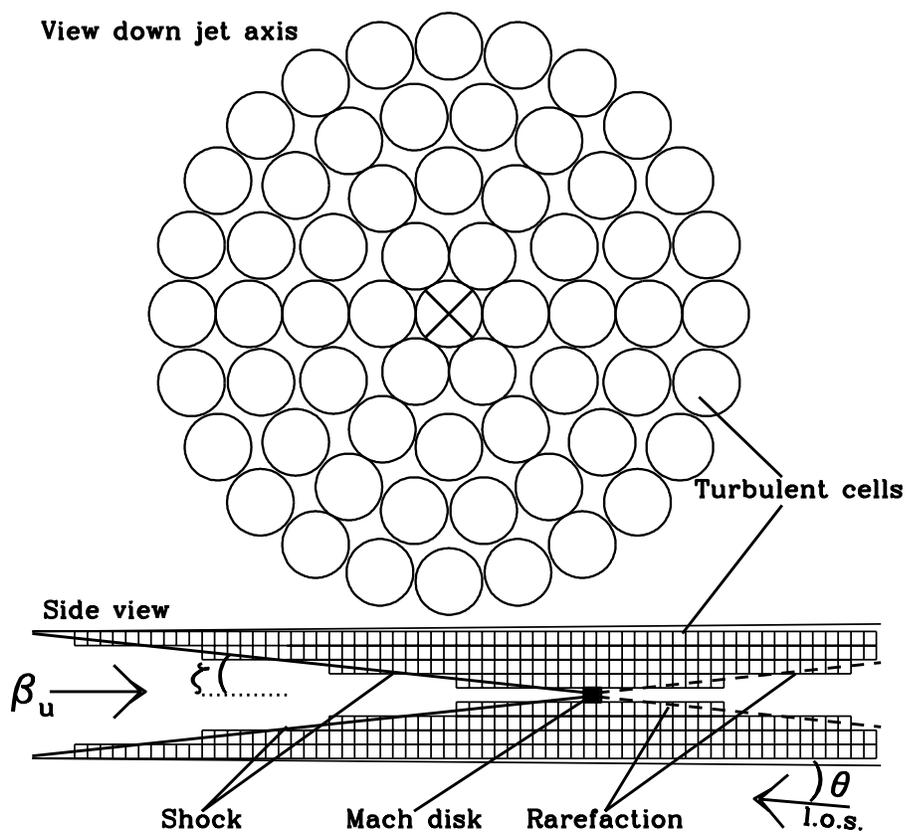}
\caption{Sketch of the geometry employed to carry out calculations within the TEMZ code.
The number of fixed computational cells across the jet cross-section (view down axis,
in which $\times$ marks the Mach disk) can be as high as 1140, not including the Mach disk.
A turbulent cell of plasma, moving at laminar velocity $\mathbf \beta_u$ upstream of the shock and
at laminar velocity $\mathbf \beta_d$ after it passes the shock,
crosses one computational cell during each time step.
The emission occurs between the conical standing shock and the rarefaction. The entire
region sketched lies $\gtrsim 1$ pc from the central engine.}
\label{fig1}
\end{figure}

As depicted in Figure \ref{fig1},
the TEMZ geometry consists of many cylindrical {\it computational} cells, fixed in space,
through each of which a single {\it turbulent} cell of plasma passes during one time step.
The plasma contains magnetic field and relativistic 
electrons (and positrons, if any), as well as protons that are presumed not to contribute a
significant amount of radiation. The field is uniform within
any given turbulent cell. The plasma flows downstream, i.e., radially away from the central engine,
through the computational cells at a laminar relativistic velocity ${\mathbf \beta}_{\mathbf d}c$,
where $c$ is the speed of light and the subscript ``d'' refers to the region downstream
of the shock. The corresponding bulk Lorentz factor is defined as $\Gamma_d \equiv (1-\beta_d^2)^{-1/2}$.
The plasma within each turbulent cell has a turbulent component ${\mathbf \beta}_{\mathbf t}c$
of the bulk velocity, with a randomly selected direction relative to the
systematic flow. For simplicity, and because of the high Lorentz factor at which the
fluid is advected downstream, the turbulent component of the bulk velocity is subsequently
held constant. Also for simplicity, the model approximates that there is no physical
interaction between adjacent cells.

The radiating electrons (as well as any positrons) are accelerated to relativistic energies as
the plasma crosses a conical ``recollimation'' shock, oriented such that it narrows
with distance from the central engine. A Mach disk can be present at the apex of the conical
shock. Beyond the apex,
the flow crosses a conical rarefaction that causes the flow to expand and accelerate.
The current version of the code assumes that the emission turns off after the plasma
crosses the rarefaction. While this may not be the case at lower frequencies given the long
time-scales for energy losses by the radiating electrons, the assumption lowers the computing time
considerably. Hence, the accuracy of the results presented here may decrease
toward centimeter wavelengths, where the flux
density is generally higher than that calculated in the TEMZ code. In many blazars,
the flux at such wavelengths is dominated by features well downstream of the ``core''
structure treated by the TEMZ model.

Figure \ref{fig1} sketches the geometry, as viewed both down the axis and from the side, of the
section of the jet over which the calculations are performed. (Note that the geometry of the
computational cell structure is approximated to be cylindrical. This ignores the slight
spreading of the jet expected from the small transverse components of both the laminar and 
turbulent velocities. These components are included in the calculation of Doppler factors
described below, however.) The time step, in the observer's frame, is selected as the time
required for a turbulent cell of plasma
to move downstream by one computational cell length $\ell = 0.2R_{\rm cell}/\tan\;\zeta$,
where the cell cross-sectional radius $R_{\rm cell}$ is selected by the user and $\zeta$,
also a free parameter (although subject to constraints; see below), is the angle between
the conical shock front and the jet axis. The factor of $0.2$ corresponds to 2 divided
by the number of cells (10) in an outer column (where the ``columns'' are parallel to the
jet axis) between the start of that column and the start of the neighboring column
(see Fig.\ \ref{fig1}).

The energy density of relativistic electrons $u_e$ in the upstream, unshocked plasma
fluctuates about a nominal value $u^*_e$ set by the user, modulated randomly with time within
a distribution that reproduces a power spectral density (PSD) of flux variations
with a power-law shape.  A subroutine supplied by R. Chatterjee \citep[see][]{chat08}, based
on the algorithm of \citet{tk95}, produces $2^n$ random values between $-1$ and 1 that
correspond to a power-law PSD with a slope provided by the user, where $n$ is an integer. Following
a suggestion by \citet{UMV05}, the TEMZ code exponentiates each number produced by the subroutine
in order to derive the multiplicative factor that is applied to the input value of the
energy density. \citep[A new, improved prescription by][for producing the fluctuations will be
incorporated in the future.]{EMP13} This is done for $2^{17} = 131,072$ times, which is
sufficient to allow calculations over the desired number of time steps included in the light
curves while taking into account light-travel delays across the grid of cells. The code
averages the fluctuations that are thereby produced over
ten time steps, an action that is necessary because of the discreteness of the cell
arrangement (see the previous paragraph). [The current version of the code
includes gradients in unshocked energy density only in the longitudinal direction (through the
time dependence), not in the transverse direction.] For the same reason, the orientation of the 
unshocked magnetic field vector is selected at random every ten time steps, i.e., for every
tenth turbulent cell along a given column. Within the intermediate
cells, the unit vector defining the direction of the field is rotated smoothly between the
previous and next randomly selected directions, along either the longest or shortest path
of rotation (selected randomly with 50\% probability for each). This procedure smooths
the discrete changes in field direction and prevents unphysical discontinuities in the magnetic field
vector along any given column. (There is, however, no relation between the directions
of the unshocked magnetic field lines in adjacent cells of different columns in the current version
of the code.) The magnetic field strength $B$ is determined via an assumption that the upstream magnetic
energy density is a constant fraction $f_{\rm B}$ of the nominal electron energy density $u^*_e$,
\begin{equation}
u_B = B^2/(8\pi) = f_B u^*_e.
\end{equation}
(The code contains an option, not implemented in the calculations presented here, that allows
the magnetic energy density to fluctuate with time such that it is always proportional to $u_e$.)
All energy densities, magnetic fields, electron energies, and electron energy distributions
are evaluated in the plasma rest frame unless otherwise indicated.

At every time step, electrons are injected into the cells immediately beyond the shock.
The injection is in the form of
a power-law energy distribution with slope $-p$ and low-energy cutoff $\gamma_{\rm 0,min}$, both
chosen by the user. The high-energy cutoff $\gamma_{\rm 0,max}$ falls within a range from
$\gamma_{\rm max,low}^*$ to $\gamma_{\rm max,high}^*$ also
set by the user, but with a value within this range that is either: (1) equal to
$\gamma_{\rm max,high}^*(B_\parallel/B)^2$ (but not less than $\gamma_{\rm max,low}^*$),
where the subscript $\parallel$ indicates the component of the magnetic field that is
parallel to the shock normal; or (2) selected at random within a power-law distribution (with 
slope as an input parameter). The first option, selected for the calculations presented below,
is meant to represent in a crude manner the higher efficiency of particle acceleration
by shocks when $\mathbf B$ is nearly parallel to the shock normal and when plasma turbulence is mild. 
The latter means that the scattering length of electron motions is too long for particles to
cross the shock front many times when the field is more inclined to the shock normal.
\citep[See, e.g.,][for simulations and a discussion related to this.]{sb12}

The compression ratio $\eta$ of the conical standing shock follows the expression given by
\citet{cc90}, which is derived from \citet{lb85}:
\begin{equation}
\eta = \Gamma_u\beta_u\sin\;\zeta(8\beta_u^2\sin^2\;\zeta\,-\,
\Gamma_u^{-2})^{1/2}(1-\beta_u^2\cos^2\;\zeta)^{-1/2}.
\end{equation}
Here, $\mathbf \beta_u$ and $\Gamma_u$ are the speed in light units and Lorentz factor, respectively,
of the bulk flow upstream of the shock,
and $\eta$ is defined as the ratio of the density of the shocked to
unshocked plasma. \citep[Note that the latter is the inverse of the definition used by][]{cc90}.
In order to satisfy the criterion for a shock, the angle $\zeta$ that the shock front subtends
to the jet axis must satisfy the criterion $\sin\;\zeta > (\sqrt{2}\beta_u\Gamma_u)^{-1}$ for
an ultra-relativistic equation of state.
On the other hand, if the value of $\zeta$ is too large, the shock decelerates the flow
so much (the downstream velocity component parallel to the shock normal drops to $\beta_{d,\parallel} = 1/3$)
that the effects of relativistic beaming become weak, contrary to the inference that
beaming of the emission is particularly strong in blazars.

The shock compresses the component of the magnetic field that is perpendicular to the shock normal by a factor of $\eta$. The magnetic field $\mathbf B$ is calculated in the plasma rest frame, which changes after the plasma crosses the shock. In order to calculate the downstream value of $\mathbf B$, the code transforms the upstream field to the standing shock frame (which is the same as the rest frame of the host galaxy), applies the compression factor to the component perpendicular to the shock normal, and then transforms the field to the rest frame of the downstream plasma \citep[see][]{lyut03}.

\subsection{Evolution of the Electron Energy Distribution Downstream of the Shock}

After the power-law injection of relativistic electrons as plasma crosses the shock front,
the code follows the evolution of the energy distribution as the electrons lose energy
via radiative losses. The jet opening angle $\phi$ is assumed to be so small
\citep[see][]{jor07,push12} that expansion cooling
is not important within the section of the jet under consideration. The code employs the
standard formula \citep[e.g.,][]{pach70} for the decrease of energy (in rest-mass units)
$\gamma$ with time:
\begin{equation}
\dot{\gamma} = -k_r (B^2+8\pi u_{\rm ph}) \gamma^2,
\end{equation}
where $k_r = 1.3\times 10^{-9}$ erg$^{-1}$ s$^{-1}$ cm$^3$.
Here, $u_{\rm ph}$ is the energy density of seed photons that are subject to inverse Compton 
scattering off the electrons. 
The formula assumes that scattering of the electrons off magnetic irregularities
randomizes the pitch angles many times over the energy-loss time-scale, and that
the field of seed photons is isotropic in the rest frame of the plasma. The code
calculates $u_{\rm ph}$ for each cell at each time step, but only
includes photon energies that fall below the Klein-Nishina limit, beyond which the
scattering efficiency decreases such that the energy losses are much lower than given by
the above equation. Both this approximation and the isotropy assumption lead to a modest
sacrifice in accuracy while allowing an analytical calculation of the electron energy
spectrum as a function of time, which greatly reduces the time required for computations.
The seed photons arriving at a given cell comprise both the steady source of IR emission from
a dust torus and time-variable emission from the Mach disk. Calculation of the latter is
carried out for an earlier time, when the photons arriving at the cell left the MD.

Since the plasma advects beyond the shock with a velocity $\beta_d \approx 1$, the distance
$z_{\rm loss}(\gamma)$ beyond the
shock that contains electrons of energy $\gamma$ is inversely related to $\gamma$:
\begin{equation}
z_{\rm loss}(\gamma) = (7.5~{\rm erg~cm^{-3}})(B^2+8\pi u_{\rm ph})^{-1}
[\gamma^{-1}-\gamma_{\rm 0,max}^{-1}]\Gamma_d ~{\rm pc},
\end{equation}
where the last term accounts for length contraction in the plasma frame \citep{MG85}. This equation can
be inverted to find the maximum energy $\gamma_{\rm max}(z)$
retained by electrons after they cross a distance $(z-z_s)_{\rm pc}$ beyond the shock front,
which is located at longitudinal position $z_s$:
\begin{equation}
\gamma_{\rm max}(z) = {{\gamma_{\rm 0,max}}\over{1 + (0.13~{\rm erg^{-1}~cm^{3}})
(B^2+8\pi u_{\rm ph})(z-z_s)_{\rm pc} \Gamma_d^{-1} \gamma_{\rm 0,max}}}.
\end{equation}
A similar equation for the lowest electron energy replaces $\gamma_{\rm max}(z)$
by $\gamma_{\rm min}(z)$ and $\gamma_{\rm 0,max}$ by $\gamma_{\rm 0,min}$.

The electron energy distribution within a cell follows the formula for steady injection
of electrons into a volume while radiative losses are operating \citep{kard62,pach70}:
\begin{equation}
N(\gamma,z) = Q_0 \gamma^{-p} \int_{0}^{t}[1 - k_r(B^2+8\pi u_{\rm ph})
t^\prime\gamma]^{p-2} dt^\prime,
\label{neq}
\end{equation}
where the injected electron energy distribution is a power-law of slope $-p$, and
$Q_0$, which is related to $u_e$, defines the injection rate.
Here, the upper limit $t$ is the lesser of $(z-z_s)(\Gamma_d\beta_d c)^{-1}$ and
$[(\gamma_{\rm 0,max}/\gamma) - 1][k_r(B^2+8\pi u_{\rm ph})\gamma_{\rm 0,max}]^{-1}$.
For a single cell with injection at one end, the integral can be evaluated as:
\begin{eqnarray}
N(\gamma,z) = Q_0 [(p-1)k_r(B^2+8\pi u_{\rm ph})]^{-1} \gamma^{-(p+1)} \times \nonumber\\
\{1-[1 - k_r(B^2+8\pi u_{\rm ph})t\gamma]^{p-1}\}.
\label{neq2}
\end{eqnarray}
This equation is valid for $\gamma_{\rm 0,min} \leq \gamma \leq \gamma_{\rm max}(z)$ and only
for plasma in computational cells that border the standing shock. For all cells
farther downstream than this, the code takes into account the change in seed photon energy density
with time in a given turbulent cell of plasma as it propagates through the computational cells
downstream of the shock. That is, the value of $u_{\rm ph}$ used in equation (\ref{neq2}) is
the average since the time that the plasma crossed the shock.

In the MD, where the magnetic field and density of electrons are much higher than in
the plasma that crosses the conical shock, the electrons are in the ``fast cooling regime,''
i.e., they radiate most of their energy before crossing one cell length. The ratio of the
photon to magnetic energy density adopted to compute the energy losses of electrons in the MD
plasma follows that calculated by \citet{se01}, $u_{\rm ph} = \varepsilon B^2/(8\pi)$,
where $\varepsilon = 0.5[\sqrt{1+(4/f_B)}-1]$.

\subsection{Calculation of Synchrotron Emission and Absorption}

The code computes the synchrotron emission coefficient $j_\nu$ within each cell at every time step via
the standard formula \citep[e.g.,][]{pach70}
\begin{equation}
j^S_\nu(\nu^\prime) = {{\sqrt{3}e^3}\over{4\pi mc^2}}B\sin\;\psi
\int_{\gamma_{\rm min}(z)}^{\gamma_{\rm max}(z)}
N(\gamma,z) {\mathcal{F}}(\nu^\prime/\nu_c) d\gamma.
\end{equation}
Here, $\nu^\prime$ is the frequency of the emission in the plasma rest frame,
$e$ is the magnitude of the electron charge, $m$ is the electron rest mass,
$\psi$ is the angle between the magnetic field and the (aberrated) line of sight in
the plasma frame, and $\nu_c = 2eB\gamma^2/(4\pi mc)$ is the synchrotron critical frequency,
averaged over electron pitch angle. The function ${\mathcal{F}}(\nu^\prime/\nu_c)$ is the usual
synchrotron kernel, given in \citet{pach70}, for example. The code uses the excellent
approximation for ${\mathcal{F}}(\nu^\prime/\nu_c)$ found by \citet{jb11}.

The angle $\psi$ is computed by Lorentz transforming the line-of-sight unit vector $\mathbf {\hat{s}}$
to the plasma (primed) frame. The coordinates are selected such that $\mathbf {\hat{s}}$ lies in the
$\mathbf {\hat{x}}$-$\mathbf {\hat{z}}$ plane. The value of $B\sin\;\psi$ is then calculated as
$|{\mathbf {\hat{s}}^\prime}{\mathbf \times B}|$.

The TEMZ code calculates the absorption coefficient in each cell in a similar manner, again
using a standard expression \citep[e.g.,][]{pach70}:
\begin{equation}
\kappa^S_\nu(\nu^\prime) = {{(p+2)\sqrt{3}e^3}\over{4\pi mc^2}}B\sin\;\psi
\int_{\gamma_{\rm min}(z)}^{\gamma_{\rm max}(z)} N(\gamma,z) \gamma^{-1}{\mathcal{F}}(\nu^\prime/\nu_c) d\gamma.
\end{equation}
The intensity of radiation leaving the cell follows the usual solution to radiation transfer in a
uniform source: $I_\nu(\nu^{\prime}) = [j_\nu(\nu^\prime)/\kappa_\nu(\nu^\prime)]
\{1-\exp[-\tau_\nu(\nu^\prime)]\}$,
where $\tau_\nu$ equals $\kappa_\nu$ times the path length through the cell, which is
equal to the cell length to a very good approximation for small angles to the line
of sight $\theta_{\rm los}$. If there are $n_{\rm ext}$
other cells along the line of sight between the cell of interest and the observer, the intensity is
attenuated by a factor of $\exp[-n_{\rm ext}\tau_{\nu}(\nu^\prime)]$. This approximation is
employed to conserve computation time. It is roughly valid since the electrons responsible
for absorption have low energies, and thus are not strongly affected by radiative energy losses.
(A more accurate calculation of absorption is under development.) The flux density from
a cell equals the intensity of radiation received from the cell times the essentially
circular area of its face. The total flux density $F_\nu$ from the source is a simple
sum over that from all cells.

\subsection{Calculation of Inverse Compton Emission}

For a given spectral intensity $I_{\nu^\prime}$ of seed photons arriving at a cell, the emission
coefficient of IC scattering requires a double integration over electron energy
and seed photon frequency. While it also properly involves a double integral over photon angles,
this requires excessive computing time when many emission zones are involved. The code therefore
utilizes the Klein-Nishina cross-section derived by \citet{dm09} (eq. 6.31) under the ``head-on''
approximation that the photon is scattered in the initial direction of the electron's motion:
\begin{equation}
\sigma_{\rm KN} = {3\over{8}}\sigma_T(y^{\prime}+y^{{\prime}\;{-1}} - 2q^{\prime} + 
q^{{\prime}\;2})/(\gamma\epsilon_i^{\prime}),
\label{knx}
\end{equation}
where $y^{\prime} \equiv 1-(\epsilon^{\prime}/\gamma)$, $q^{\prime} \equiv 
\epsilon^{\prime}/(\gamma y^{\prime} \epsilon_i^{\prime})$, and
$\epsilon^{\prime} \equiv h\nu^{\prime}/(mc^2)$, $\nu_i^{\prime}$ is the frequency of the
seed photon, $\nu^{\prime}$ is the
frequency of the scattered photon, $\sigma_T = 6.653\times 10^{-25}$ cm$^2$ is the Thomson
cross-section, and $h$ is Planck's constant. The cross-section is valid for
$(1+4\gamma\epsilon_i^{\prime})^{-1} < y^{\prime} < 1$.

The inverse Compton emission coefficient is then
\begin{equation}
j^{IC}_\nu(\nu^\prime) = \int_{\gamma_1}^{\gamma_{\rm max}}\int_{\nu_{i,1}^{\prime}}^{\infty}
I^\prime_{\rm seed}(\nu_i^{\prime}) N(\gamma,z) \sigma_{\rm KN}(\nu^{\prime},\nu_i^{\prime},\gamma)
d\nu_i^{\prime} d\gamma,
\end{equation}
where $\gamma_1$ is the greater of $\gamma_{\rm min}$ and
$\epsilon^{\prime}[1+\sqrt{1+(\epsilon_i^{\prime}\epsilon^{\prime})^{-1}}]/2$,
$\nu_{i,1}^{\prime} = \nu^{\prime}[4\gamma^2(1-\epsilon^{\prime}\gamma^{-1})]^{-1}$, and
$I^\prime_{\rm seed}(\nu_i^{\prime})$ is the spectral intensity of seed
photon emission as measured in the plasma rest frame.

Equation (\ref{knx}) contains an implicit dependence on the angle between the bulk velocity vector
of the plasma and the direction of the source of seed photons. This results from the
Doppler formula for the transformation of frequency from the rest frame of the seed photon
source (SP) to the plasma frame: $\nu_i^{\prime} = \nu_i\delta_{\rm SP}$, where
\begin{equation}
\delta_{\rm SP} = [\Gamma_{\rm rel}(1-\beta_{\rm rel}\cos\;\vartheta^\prime_{\rm rel}]^{-1}.
\end{equation}
Here, $\beta_{\rm rel}$ is the magnitude of the relative velocity in light units between
the plasma and seed photon source, $\Gamma_{\rm rel}$ is the corresponding Lorentz factor,
and $\vartheta^\prime_{\rm rel}$ is the aberrated angle between the velocity vector of the plasma
and the aberrated direction of the seed photon source in the plasma frame.

\subsection{Sources of Seed Photons for Inverse Compton Scattering}

The current version of the TEMZ code includes two sources of seed photons available for IC
scattering by electrons in the cells: infrared (IR) blackbody emission from hot dust
contained in a patchy molecular torus (external Compton, hereafter ``EC-Dust'') and synchrotron plus
first-order synchrotron self-Compton emission from a Mach disk (``SSC-MD'')
centered on the jet axis at the narrow end of the conical shock
(see Fig.\ \ref{fig1}). Because of the relativistic motion of the plasma in the cells, the
radiation from both of these sources is blueshifted and beamed in the plasma rest frame.
The code does not currently include the seed photons from synchrotron emission within the cell
under consideration (``SSC'') plus all of the other cells (``SSC-C''),
owing to the excessive computational time
required given the large number of cells. Future parallelization of the C++ version of the
code is planned in an effort to make such calculations possible.

The central circle of the dusty torus lies a distance $r_t$ from the center of the system, assumed to
be a black hole (BH), and the torus has a cross-sectional radius $R_t$. As viewed from a cell at a polar
distance $z$ from the BH, the dusty torus subtends a ring in the sky. In the rest frame of the
system, the direction of the outer/inner boundary of the ring subtends an angle
\begin{equation}
\xi^{\rm out}_{\rm in} \approx \tan^{-1}\;(r_t/z)\,{\pm}\, \sin^{-1} [R_t(z^2+r_t^2)^{-1/2}]
\end{equation}
to the direction toward the BH.
In the rest frame (primed) of the plasma in the cell, which moves at velocity ${\mathbf \beta}_{\mathbf d}$
(Lorentz factor $\Gamma_d$) approximately in the ${\mathbf {\hat{z}}}$ direction (the slight deviation
from this direction is unimportant for this calculation), each of these angles is transformed as
\begin{equation}
\tan\;(\xi^\prime/2) = \tan\;(\xi/2)[\Gamma_d(1+\beta_d)]^{-1}.
\end{equation}
The blackbody emission from the torus in the plasma frame corresponds to a temperature
\begin{equation}
T^\prime = \delta_{\rm SP}T = [\Gamma_d(1-\beta_d\cos\;\xi^\prime)]^{-1}T,
\end{equation}
where $\xi^\prime$ ranges from $\xi^\prime_{\rm in}$ to $\xi^\prime_{\rm out}$.
The code integrates the resultant blackbody spectrum over the surface area from angle
$\xi_{\rm in}$ to $\xi_{\rm out}$.
The area filling factor of the blackbody emission is set by the size, temperature, and luminosity
of the torus, with the latter derived from the IR flux observed at the Earth. Direct observations
of the dust emission from the quasar 4C~21.35 \citep{malm11} guide the choice of
these physical parameters. In this $\gamma$-ray bright object, the flux of the dust emission is
almost entirely from a single-temperature blackbody ($T \approx 1200$ K) and has a value
of 0.22 times the flux of the accretion disk that provides most of the optical-ultraviolet
luminosity. In the case of 4C~21.35, the area filling factor of the dust emission is
$\sim 0.6~{\rm pc}^2/(r_t R_t)$ and $R_t \approx 0.22r_t$.

The code calculates the synchrotron and first-order SSC emission generated by the plasma in
the MD at each time step (see above). Since the cross-sectional size of the MD is not
well specified by gas dynamical theory, it is left as a free parameter (see Table \ref{tab1}).
The plasma decelerates from a high Lorentz factor to $\Gamma_d \approx 1.2$ as it crosses the MD
shock front, hence the shock greatly magnifies the magnetic field and electron density
relative to the values in other cells. The seed photons from the MD arrive at each cell
with a time delay that depends on the distance of the cell from the MD. The blueshift
$\delta_{\rm MD}$ of the MD photons in the frame of the plasma in the cell depends on the
angle between the cell's line of sight to the MD and the velocity vector of the plasma relative
to that of the Mach disk plasma, with the latter assumed to be $\beta_{\rm MD} = 1/3$ in
the ${\mathbf {\hat{z}}}$ direction.
The flux then depends on both the inverse square of the distance of the cell from the MD and
the relativistic beaming factor $\delta_{\rm MD}^{2+\alpha}$, where the dependence of the
flux density $F_\nu$ on frequency $\nu$ is $F_\nu \propto \nu^{-\alpha}$. (Since $\alpha$ can
change with frequency, its value is determined by the code from the emission calculation.)
The synchrotron
flux received from the MD also depends on the angle $\psi_{\rm MD}$ that the MD magnetic field
subtends to the aberrated line of sight between the cell and the MD,
$F_\nu \propto (\sin\;\psi_{\rm MD})^{1+\alpha}$. The synchrotron self-absorption optical depth
within the MD for radiation that will intersect the cell is proportional to
$(\sin\;\psi_{\rm MD})^{1.5+\alpha}$. The current version of the code approximates the optical depth of synchrotron self-absorption in intervening cells by multiplying the absorption coefficient of 
the cell by the distance between the MD and the cell.

\subsection{Linear Polarization}

The current version of the TEMZ code calculates the linear polarization for each cell in a simple manner. For an optically thin cell, the degree of polarization depends on the (frequency-dependent) spectral index, $\Pi_{\rm cell} = (\alpha+1.0)/[\alpha+(5.0/3.0)]$,
with electric-vector position angle (EVPA) perpendicular to the direction of the magnetic field as projected on the sky. In the optically thick case $\Pi_{\rm cell} = 3/(12\alpha+19)$, and the EVPA is along the projected direction of the magnetic field \citep{pach70}. These relations are valid because the magnetic field is treated as uniform within each cell. The code calculates the value of $\chi_{\rm cell}$, aberrated by relativistic motion, following the formulation of \citet{lyut05}. From the flux
density of the cell $F_{\nu,{\rm cell}}$ determined as in {\S}2.2, the Stokes
parameters are then set as $Q_{s,{\rm cell}} = F_{\nu,{\rm cell}} \cos\;2\chi_{\rm cell}$
and $U_{s,{\rm cell}} = F_{\nu,{\rm cell}} \sin\;2\chi_{\rm cell}$.
The code sums $Q_{s,{\rm cell}}$ and $U_{s,{\rm cell}}$ over all of the cells to determine the
integrated degree of polarization $\Pi = (Q_s^2+U_s^2)^{1/2}/F_\nu$ and position angle
$\chi = 0.5\tan^{-1}(U_s/Q_s)$.

\section{Testing of the Numerical Code}

Because of the complexity of the algorithms needed to carry out the calculations, the author has tested the output of the code in a variety of ways. The first set of tests involved comparison of the synchrotron and IC spectrum computed for single cells, with standard analytical expressions for a cylindrical source with uniform properties \citep[see, e.g.,][]{pach70}. The position angle of linear polarization of the synchrotron emission was tested for different directions of uniform magnetic fields relative to the shock front and to the line of sight. The difference in degree and position angle of polarization with shock compression was compared with the case of no compression. A print-out of the evolution of the electron energy distribution of a given parcel of plasma provided a check on the numerical handling of how radiative energy losses affect the distribution. The code passed all of these basic tests.

\begin{figure}
\epsscale{1.0}
\vspace{-3cm}
\plotone{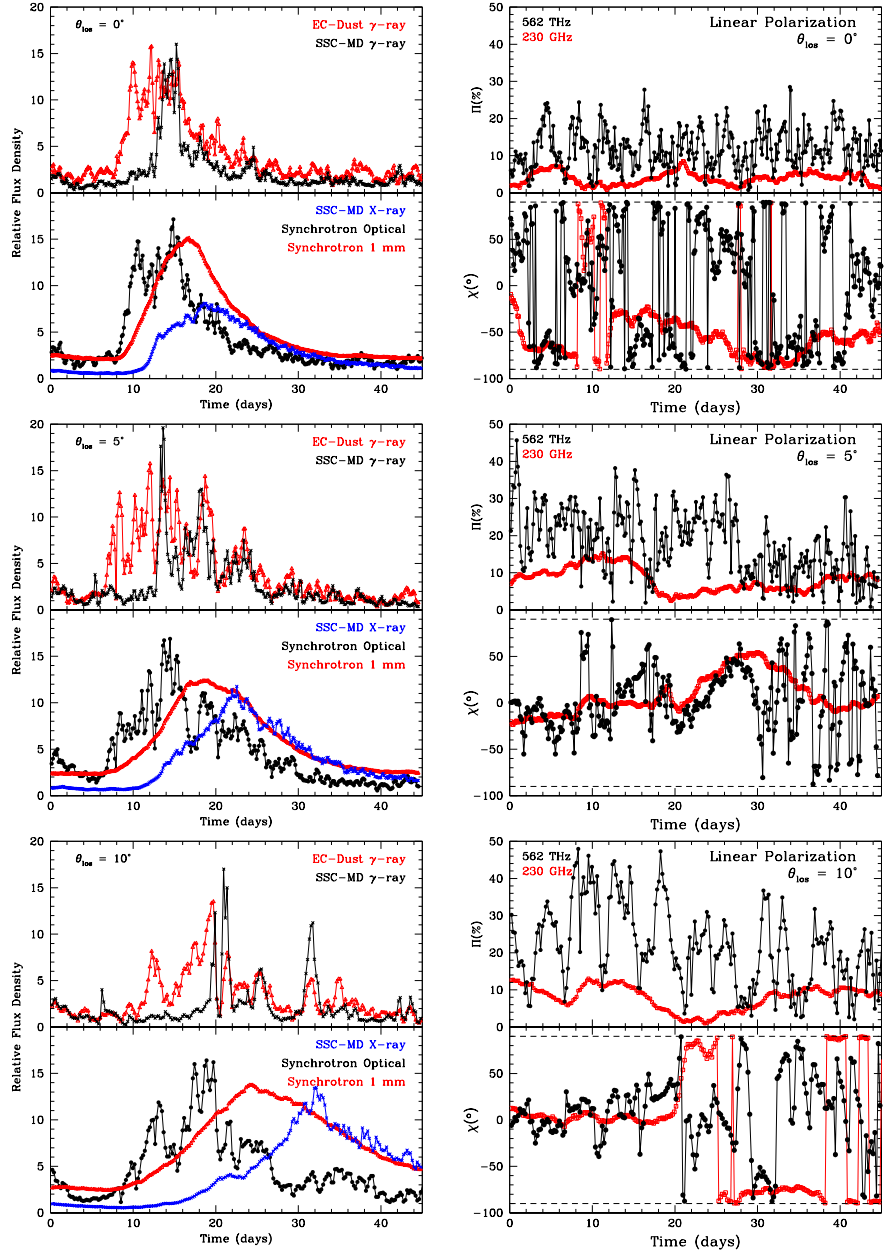}
\vspace{-4cm}
\caption{Light curves {\it(left)} and polarization vs. time {\it (right)} simulated with the TEMZ code
for the test case of a sudden, short increase in the injection rate of relativistic electrons, as described in the
text. The overall profiles reflect the effect of the chosen geometry (see Fig.\ \ref{fig1}) plus, on
short time-scales, the effect of turbulence. The three cases displayed correspond to different
viewing angles $\theta_{\rm los}$. Note that polarization position angle $\chi = +90^\circ$ is the
same as $-90^\circ$. The projected position angle of the jet axis is $0^\circ$.}
\label{test}
\end{figure}

Testing of the full multi-zone code involved checks on the various time delays for propagation of electromagnetic waves and advection of plasma downstream of the shock. The introduction of a very high-amplitude pulse of relativistic electron injection (i.e., a sudden increase in $Q_0$) over a few time steps facilitated comparison of the time-dependent output of the code with the expected behavior. Sample results are displayed in Figure \ref{test}. Here we see that, for $\theta_{\rm los} \approx 0$, the induced optical synchrotron and EC-Dust flux rises steeply as the pulse crosses the ring of computational cells at the upstream end of the shock, then declines more gradually as the pulse crosses smaller rings of cells. (The electrons that produce $\gamma$-ray and optical photons lose energy rapidly, hence radiation at these frequencies is confined to a thin layer close to the shock.) The SSC-MD flux increases modestly at first before a rapid flare occurs as the pulse crosses the MD, creating a sudden increase in seed photon flux.

\section{Results of Sample Numerical Simulations}

Figure \ref{sims} presents samples of light curves, polarization vs.\ time, and SEDs for two of the
TEMZ simulations. The parameters were selected to be similar to those of two blazars,
BL {\it Lacertae} and PKS~1510$-$089, as derived from observations \citep[e.g.,][]{jor05,mar08,mar10}.
They are not intended, however, as actual fits to any data. In fact, because of the randomness
inherent in the model, close fits to observational data cannot be attained. Rather, future success
of the model will depend on comparison of the statistical characteristics of the simulated results
with those of the data.

The BL Lac-like simulation bears a resemblance to the observed flux and polarization versus time curves of this
blazar presented in \citet{mar13}. The running mean of the $\gamma$-ray flux level increases with time,
while on short time-scales the flux varies erratically. The fluxes at lower frequencies rise and fall
roughly together, although with some cross-frequency delays in maxima and minima. The fluctuations
about the running mean at optical and X-ray frequencies occur on similar time-scales as at $\gamma$-ray
energies, but with lower amplitudes, while the variations at $\lambda$1 mm (230 GHz) are quite smooth
owing to the larger volume occupied by electrons that emit at this wavelength relative to those
responsible for the optical emission.
There are ``orphan'' (i.e., with no counterpart at the other wavebands) $\gamma$-ray flares (e.g.,
at $t \sim 30$ days) and optical flares without $\gamma$-ray counterparts (e.g., at $t \sim 46$ days).
The higher amplitude of the $\gamma$-ray variations relative to those at optical frequencies results
both from the somewhat higher (factor of $\sim 2$-3) energies --- and therefore shorter radiative lifetimes --- of electrons that produce $\gamma$-rays than those that produce optical synchrotron
radiation, as well as variations in seed photon
energy density throughout the emission region owing to changing conditions in the MD.
Since IR emission by hot dust has yet to be confirmed in a BL Lac object, the simulation
includes no contribution to the seed photon field from this component.

The simulation of a blazar similar to the quasar PKS~1510$-$089 (bottom row of Fig.\ \ref{sims}) produces
SSC-MD $\gamma$-ray and synchrotron optical light curves that are quite similar, although not
identical, in profile. The EC-Dust $\gamma$-ray flux varies more erratically because the electrons
that scatter the dust-emitted photons to $\gamma$-ray energies are near the highest energies
generated by the shock, and therefore occupy a small fraction of the total volume. Such high energies
result from the lower mean frequency of the seed photons from the dusty torus relative to the synchrotron
and SSC emission from the MD. [Note that the calculation of the seed photon field includes
integration over the torus (see {\S}2.4), rather than the crude (unless the opening angle of the
torus is very large) approximation that the dust-emitted
photons are isotropic in the quasar rest frame at the position of the jet plasma, as approximated,
for example, by \citet{gt09}.] The X-ray and $\lambda$1 mm light curves are well-correlated, as
observed \citep[e.g.,][]{mar10}, although they do not follow each other completely.

Many of the flares have quite sharp peaks, with rises and decays that are approximately linear
or exponential. This is a common characteristic of blazar light curves \citep[e.g.,][]{chat12}
that is difficult for many models to reproduce. The time-scales of the variations, especially
at optical and $\gamma$-ray frequencies, can be extremely short, as anticipated by \citet{mj10}
and \citet{np12}. This is a result of the combination of (1) the small volume filling factor of 
production of the highest energy electrons, (2) the short radiative lifetimes of these electrons,
(3) the high Lorentz factor of the laminar flow, and (4) the turbulent peculiar velocities
$\mathbf \beta_t$, which can increase the Doppler factor by as much as a factor of $(1-\beta_t^2)^{-1/2}$.

The dependence of the volume filling factor on frequency leads to two additional features of
the model that compare well with observations \citep[see][]{jor07,mj10,W12,jor13}. The time-averaged
degree of linear polarization increases, and the time-scale of variability of
both the degree and position angle of polarization decreases while the amplitude of variability increases, with frequency. Furthermore,
breaks in the spectrum by more than by 0.5 \citep[expected from radiative losses in a
single emission zone; e.g.,][]{MG85} appear at IR and $\gamma$-ray frequencies.

As with the light curves, the polarization versus time curves in Figure \ref{sims}
(middle panels) also exhibit a
combination of systematic and erratic behavior. As expected, the optical polarization (represented
in the figure by 562 THz, or 534 nm, which is within the $V$ band) ranges from
close to zero up to tens of percent for the BL Lac-like simulation and up to nearly 20\%
for the quasar-like simulation, similar to the observed range
\citep[see][to compare with data]{mar10,mar13}. The mean EVPA
fluctuates about the jet direction ($0^\circ$ in the simulations), as is generally observed
in BL Lac. Apparent rotations of the polarization vector occur, although a larger number of
simulations is needed to determine whether they resemble the variety of such rotations that
are observed in blazars \citep[e.g.,][]{mar08,Lar08,abdo10,Lar13}. Such rotations are common
when the degree of polarization is low \citep[see also][]{jones88,MGT92}. If the
randomness in the magnetic field directions causes the polarization integrated over
$\mathcal{N}-\mathcal{M}$ of the cells to cancel, with $\mathcal{N} \gg \mathcal{M}$,
the net polarization is low and determined by
the other $\mathcal{M}$ cells. Since a small number of cells is involved, time variations
can occur that resemble rotations of the polarization vector. However, such apparent
rotations are unlikely to explain the systematic optical EVPA rotations reported in
BL Lac \citep{mar08} and PKS~1510-089 \citep{mar10}. Both of these observed events
occurred prior to the time when the emitting plasma crossed the core, and hence took
place upstream of the location where the TEMZ calculations apply. Despite the slower
variations of polarization at longer wavelengths, rotations of the polarization vector
still occur at millimeter wavelengths both in the simulations and in observations
\citep[e.g.,][]{Lar08}.

Both the degree and position
angle of the quasar-like simulation vary more erratically than is typically observed in
PKS~1510$-$089 \citep[e.g.,][]{mar10}. This discrepancy is likely the result of the
lack of continuity in the magnetic field vector across cells in the direction transverse to
the jet axis in the current version of the TEMZ code (see {\S}2). Actual jets may
also contain an ordered component of the magnetic field in addition to the turbulent
component. If so, the very short time-scale variations in polarization will have lower
amplitudes.

Figure \ref{images} presents sample total ($I$) and polarized ($P$) intensity images at
43 GHz generated
at a particular time of the quasar-like simulation. These images serve as representations
of the brightness and polarization distribution across the source, as well as an indication
of the ability of VLBI observations to test the TEMZ model. In order to generate the images,
a rectangular grid of pixels is set up, with the intensity from all of the cells that lie along
the line of sight represented by each pixel summed (vectorially for a $P$ image) to obtain the
intensity of the pixel. The pixelated image is then convolved with a circular Gaussian restoring
beam (i.e., a point-spread function). The image at the top of the figure is convolved with a
small beam to reveal the underlying intensity pattern, while the image at the bottom has
a FWHM resolution equal to the diameter of the jet cross-section. The latter is typical of
the angular resolution of the longest baselines of a millimeter-wave VLBI observation with
nearly Earth-diameter maximum separations of antennas. The addition of one or more space-based
antennas is therefore needed in order to determine whether a given blazar possesses the brightness
and polarization structure predicted by the model.

Comparison of the top right and bottom right images in Figure \ref{images} demonstrates that
the low net polarization at 43 GHz is caused by cancellation of cross-polarized sections,
which is nearly complete for very small viewing angles because of the underlying symmetry
of the conical shock structure. The deviation from perfect symmetry between the top and
bottom halves of the images is caused by the random elements in the TEMZ model.
Except for this effect, which is small at 43 GHz owing to the large number of cells that
emit at this frequency, the polarized intensity structure is similar to that calculated and
displayed by \citet{caw06}, \citet{nal09}, and \citet{cjm13}. These authors consider the
somewhat different case of a standing shock compressing a magnetic field that is completely
chaotic upstream of the shock.

\section{Conclusions}

\begin{figure}
\epsscale{1.2}
\vspace{-6cm}
\plotone{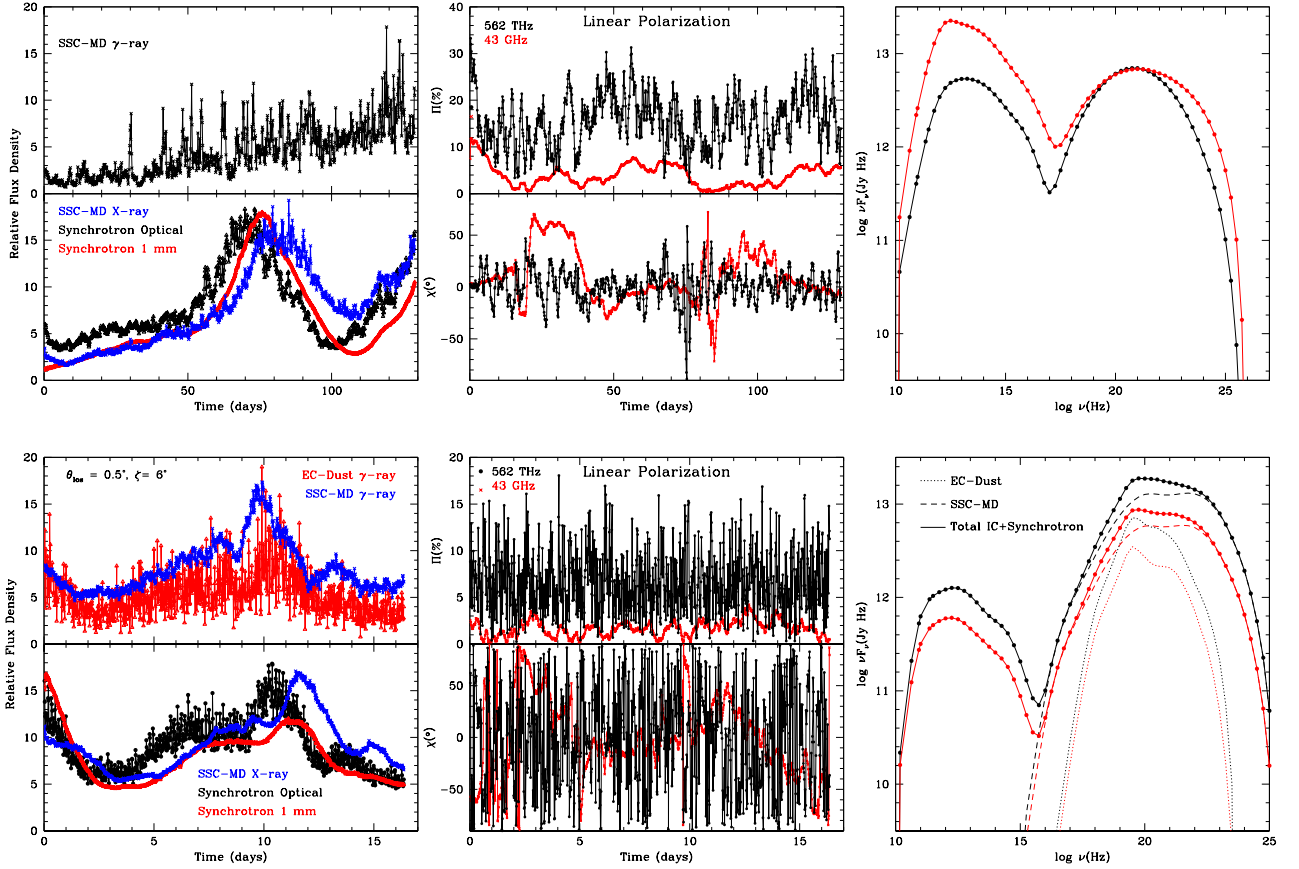}
\vspace{-10cm}
\caption{Sample light curves {\it (left)}, polarization vs. time {\it (middle)}, and SED
at two times {\it (right)} from TEMZ simulations. Flux levels are scaled arbitrarily
so that the shape of each light
curve is apparent. {\it Top row:} parameters selected to be similar to BL {\it Lacertae}:
$n_{\rm rad}=7$ (168 cells across the shock front),
$z_{\rm MD}= 1.22$ pc, $Z=0.069$, $\alpha=0.55$, $b=1.7$, $B=0.04$ G, $f_B=1.0$, $R_{\rm cell}=0.003$ pc,
$\gamma_{\rm min}=300$, $\gamma_{\rm max,high}=140,000$, $\gamma_{\rm max,low}=7000$, $\beta_u=0.990$,
$\beta_t=0.577$, $\zeta=10^\circ$, $\theta_{\rm los}=7.7^\circ$, $\phi=1.9^\circ$, $A_{\rm MD}=0.01$, and
no significant thermal emission from dust. The SEDs correspond to times of 70.76 days {\it (red)} and
107.6 days {\it (black)}. Temporal resolution is 0.129 days (3.1 hr) over 1000 time steps. {\it Bottom row:} parameters selected to be similar to PKS~1510$-$089: $n_{\rm rad}=10$ (270 cells across the shock front),
$z_{\rm MD}= 1.18$ pc, $Z=0.361$, $\alpha=0.7$, $b=1.7$, $B=0.04$ G, $f_B=1.0$, $R_{\rm cell}=0.001$ pc,
$\gamma_{\rm min}=1800$, $\gamma_{\rm max,high}=37,500$, $\gamma_{\rm max,low}=5000$, $\beta_u=0.99969$,
$\beta_t=0.577$, $\zeta=6^\circ$, $\theta_{\rm los}=0.5^\circ$, $\phi=0.2^\circ$, $A_{\rm MD}=1$,
$T_{\rm dust}=1200$ K, $L_{\rm dust}=1\times10^{46}$ erg s$^{-1}$,
$r_{\rm dust}=3.0$ pc, and $R_{\rm dust}=0.8$ pc. The SEDs correspond to times of 3.0 days {\it (red)} and
10.0 days {\it (black)}. The temporal resolution is 0.0159 days (23 min) over 1027 time steps.
}
\label{sims}
\end{figure}

\begin{figure}
\epsscale{0.9}
\vspace{-3cm}
\plotone{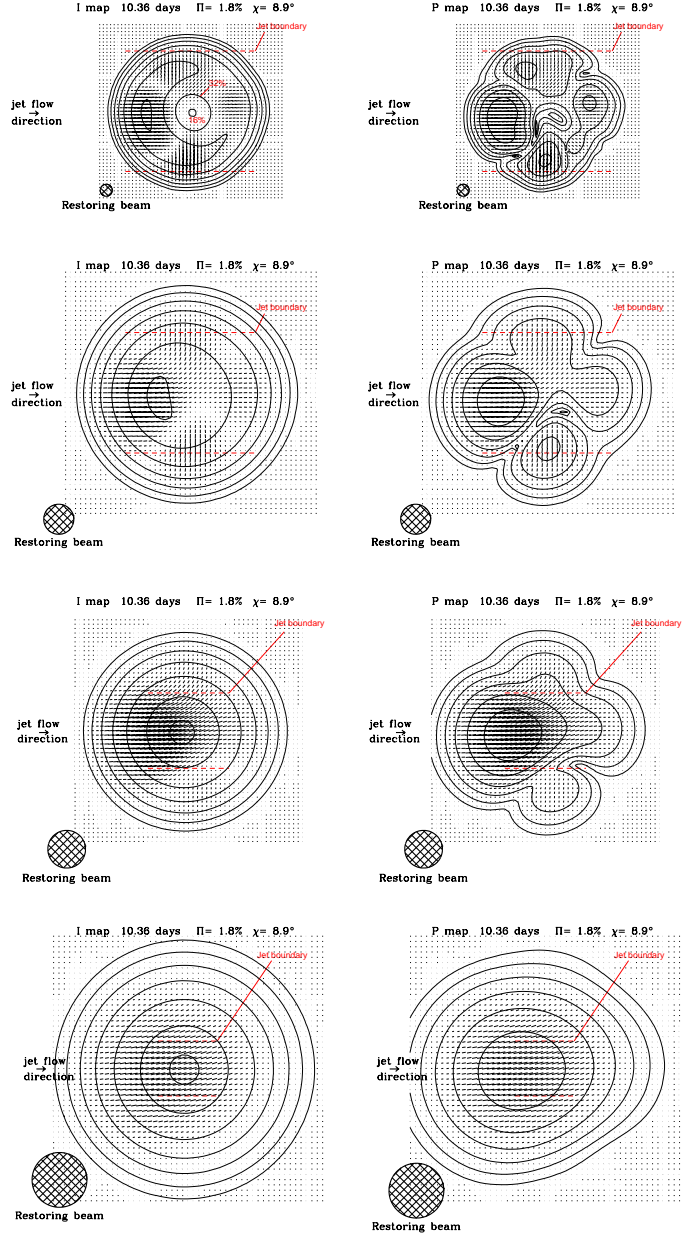}
\vspace{-1cm}
\caption{Simulated total ({\it left}) and polarized ({\it right}) intensity images at 43 GHz of the
section (denoted by dashed lines) of a blazar jet whose emission is calculated by the TEMZ code.
Contour levels correspond to 2, 4, 8, 16, 32, and 64\% of the peak intensity; for the total
intensity images, an additional 95\% contour is added to define the location of the peak. (Note
that the two contours near the center of the top left image represent decreases in intensity,
to 32\% and 16\%, as marked.)
The line segments represent the orientation of the polarization electric vector, with lengths
proportional to the polarized intensity. All images are from time 10.36 days of Figure \ref{sims}
(bottom panels). The intensity is smoothed with a circular Gaussian restoring beam
(point-spread function), with progressively decreasing FWHM diameter from top to bottom,
as indicated to the bottom left of each image.
}
\label{images}
\end{figure}

The many-zone emission model provides much more realistic emission calculations than do schemes
involving a very small number of zones. The TEMZ model, improved in ways discussed
below, therefore holds some promise in explaining the diversity of blazar variability in
multi-waveband flux, polarization, and structure.
While this may seem to be at the expense of complexity,
the number of free parameters, which are listed in Table \ref{tab1}, is similar to that of single-zone
models, with many of the parameters ($Z$, $\alpha$, $b$, $\beta_u$, $\theta_{\rm los}$, $\phi$,
$T_{\rm dust}$, and $L_{\rm dust}$) specified by observations. Furthermore, instead of fitting
the data to SEDs measured at a small number of times, the results produced by the TEMZ code
can be compared with the time evolution of the SED as well as the characteristics of the light
curves, degree and position angle of polarization versus time, and total as well as polarized
intensity images of blazars. The author plans to explore how these characteristics as produced
by the simulations depend on the adjustable parameters. This will include timing analyses of
the simulated flux and polarization curves so that the cross-frequency correlations and
PSDs can be compared with those observed in blazars. The results displayed in Figure \ref{sims}
and discussed in {\S}4 resemble observations of blazars well enough to warrant such a study.

There is, on the other hand, one aspect of the simulations that
conflicts with the data: the ratio of $\gamma$-ray to X-ray luminosity (right-most panels of
Fig.\ \ref{sims}) is not as high as observed during major outbursts of some of the most
prominent $\gamma$-ray bright quasars. This problem with the simulations was pointed out by
\citet{W12}, who used the TEMZ code to fit the SEDs of the quasar 3C~454.3 during a major
outburst. The over-production of X-rays in the model is caused by the strong synchrotron
emission from the MD at far-IR wavelengths, as viewed in the rest frame of the
plasma in the other cells. Perhaps a more refined treatment of the MD, based on
magneto-hydrodynamic (MHD) simulations (see below), will solve this problem. On the other hand, 
a thermal spectrum of seed photons, which declines sharply (as $\nu^2$) toward
low frequencies, does not produce an X-ray excess as long as $\gamma_{\rm 0,min} \gtrsim 100$.
The high $\gamma$-ray to X-ray flux ratio is one reason why seed
photons from thermal emission from broad emission-line clouds and a dusty torus is favored
by many authors \citep[e.g.,][]{gt09}. However, at distances $\gtrsim 3$ pc from the central
engine, the seed photon field from these regions is too weak to explain the high $\gamma$-ray
luminosity. If, instead, some stray molecular clouds or dense, ionized clouds
are located along the periphery of the jet, it is possible that they could produce a
sufficient density of seed photons to explain the $\gamma$-ray emission from IC
scattering \citep{LT13,isler13}. The author plans to add such a source of seed photons to
the code. Also, in order to simulate blazars with ratios of $\gamma$-ray to infrared
luminosities closer to unity, development of a full SSC and SSC-C calculation, with seed photons
from all of the cells (calculated in retarded time), is underway.

It is likely that, in many blazars, other physical processes besides a single standing
shock are responsible for energizing electrons to induce flares as well as more quiescent
emission. In fact, there is evidence in a number of blazars for multiple standing features
that could correspond to oblique or conical shocks \citep[e.g.,][]{jor05}. The TEMZ model
could be applied separately to emission from any such stationary structures, although if there is
a non-conical shock, the code would need to be altered accordingly. Moving shocks in a jet
filled with turbulent plasma \citep{MGT92} are possible as well, especially when a very bright 
superluminal
knot is observed to propagate down the jet. Magnetic reconnections have been proposed as
an alternative to shocks as a primary process for accelerating electrons, causing
relativistic bulk motions of plasma, and therefore generating flares \citep[e.g.,][]{gian09}.
The TEMZ code can potentially be modified to incorporate this scenario after studies of
relativistic reconnection elucidate how the magnetic field topography and electron energy
distribution vary with time and position.

While randomness in the magnetic field direction in different turbulent cells can cause
observed rotations in the linear polarization vector, it probably does not explain the
systematic rotations observed in some blazars. The conclusion that such events occur before
a disturbance in the jet reaches the standing shock implies that the variable emission described
in the TEMZ code does not cover all of the events in the light curves of blazars. This is
also true at lower frequencies where much of the emission comes from regions farther downstream
in the jet, and even in some cases at $\gamma$-ray energies where observations indicate
that flares can take place many parsecs from the core \citep[][]{agudo11a,agudo11b}.
Nevertheless, there is strong evidence that most of the multi-waveband outbursts in blazars are
associated with emission in the core region treated in the TEMZ model \citep{mar12}.

Despite the promising results presented here, more realistic simulations are needed
to confront the model with data in a serious way. The geometry and physical properties of
the standing shock and Mach disk system are simplified in
the current version of the TEMZ code. Ideally, they should be determined instead by relativistic MHD
simulations. The energization of relativistic electrons is also treated in an {\it ad hoc}
manner, as is the approximation to turbulence. Furthermore, the fluctuations in electron energy
density are forced to agree with the observed PSD of blazar flux variations, while they
ideally should be dictated by successful simulations of flow from the accreting black hole
system into the jet. The author plans to collaborate with
other groups working in these areas to combine the TEMZ emission code with detailed calculations of
the dynamics and plasma physics of relativistic jets in order to produce more realistic
simulations of emission from blazar jets.

\acknowledgments
The author thanks R. Chatterjee for providing his code for producing random fluctuations that
obey the statistics of a power-law PSD,
S.\ Jorstad, M.\ Joshi, J.\ L.\ G\'omez, N. MacDonald, M. Baring, and M.\ B\"ottcher for
discussions that were useful in the development of the model (and the first four plus
M. Malmrose for critical readings of a draft version of the manuscript), M. Valdez for
scrutinizing, and identifying
some flaws in, the numerical code and for translating it to C++, and J.\ L. G\'omez
for providing visualization software that aided in testing of the results. He also thanks Prof.
Stefan Wagner for his hospitality during a visit to the University of Heidelberg in 1998
when this study was first conceived.
Funding of this research included NASA grants NNX10AO59G, NNX11AQ03G, and NNX12AO79G,
and NSF grant AST-0907893. This work benefited from discussion within International Team
Collaboration 160 sponsored by the International Space Science Institute (ISSI) in
Switzerland.

\clearpage
\begin{deluxetable}{lccl}
\singlespace
\tablecolumns{4}
\tablecaption{\bf Adjustable Parameters in the TEMZ Computer Code\label{tab1}}
\tabletypesize{\scriptsize}
\rotate
\tablenum{1}
\tablehead{
\colhead{Symbol}&\colhead{Description}&\colhead{Range of Values}&\colhead{Notes}
}
\startdata
$n_{\rm rad}$&Number of cells between Mach disk \& jet boundary&4-19& Cross-section contains 3($n_{\rm rad}+1$)$n_{\rm rad}$ cells \\
$Z$&Redshift&---&From observations of blazar of interest \\
$z_{\rm MD}$&Distance of Mach disk/shock apex from central engine&1-20 pc& \\
$p$&$-1$ times slope of injected electron energy distribution&1.5--3&Single power-law assumed in absence of radiative losses \\
$-b$&Power-law slope of power spectral density&1.5-2&Used to determine modulation of energy density \\
$B$&Unshocked magnetic field&0.01-0.5&Modulated as square-root of energy density \\
$f_B$&Ratio of magnetic to relativistic electron energy density&0.01-1&Equipartition value is $\sim 1$ \\
$R_{\rm cell}$&Radius of a cylindrical cell&0.002-0.05 pc&Depends on scale size and value of $n_{\rm rad}$ \\
$\gamma_{\rm 0,min}$&Lowest energy of electrons accelerated by shock&10-2000&Low-energy cutoff of power-law injection \\
$\gamma^*_{\rm max,high}$&Highest possible value of $\gamma_{\rm 0,max}$ produced by shock&5000-$10^6$&Optimal acceleration of electrons by the shock \\
$\gamma^*_{\rm max,low}$&Lowest value of $\gamma_{\rm 0,max}$ produced by shock&1000-$10^4$&Lowest efficiency of acceleration  \\
${\mathbf \beta}_{\mathbf u}$&Bulk laminar velocity of unshocked plasma in light units&0.97-0.9998& \\
${\mathbf \beta}_{\mathbf t}$&Turbulent velocity of unshocked plasma in light units&0-0.577&Range assumes subsonic turbulence \\
$\zeta$&Angle between conical shock and jet axis&$4-10^\circ$&Must satisfy criterion
$\sin\;\zeta > (\sqrt{2}\beta_u\gamma_u)^{-1}$ \\
$\theta_{\rm los}$&Angle between jet axis and line of sight&$0-20^\circ$& \\
$\phi$&Opening semi-angle of jet&$0-3^\circ$& \\
$A_{\rm MD}$&Ratio of cross-sectional areas of Mach disk \& a cell&0.01-10& \\
$T_{\rm dust}$&Blackbody temperature of hot dust torus&1000-1200 K&Observed value in two blazars \\
$L_{\rm dust}$&Luminosity of thermal IR emission from hot dust&$1\times10^{44}-2\times10^{46}$ erg s$^{-1}$& \\
$r_{\rm dust}$&Mean distance of dust torus from black hole&1-5 pc& \\
$R_{\rm dust}$&Cross-sectional radius of dust torus&0.2-0.5$r_{\rm dust}$& \\
\enddata
\end{deluxetable}

\clearpage

\clearpage

\clearpage

\clearpage

\end{document}